\documentstyle[epsf,12pt]{article}

\newcommand{\halftext}{6.6cm}
\newcommand{\V}[1]{\mbox{\boldmath{${#1}$}}}
\newcommand{\outout}{{\mbox{out}}}
\newcommand{\inin}{{\mbox{in}}}
\newcommand{\aout}{a_{\outout}}
\newcommand{\ain}{a_{\inin}}
\newcommand{\e}{\mbox{e}}
\newcommand{\Side}{{\mbox{\small side}}}
\newcommand{\Out}{{\mbox{\small out}}}
\newcommand{\Long}{{\mbox{\small long}}}

\begin{document}
\begin{flushright}
WU-HEP-97-7 \\
TWC-97-4 \\
nucl-th/9709010 \\
September 1997
\end{flushright}
\begin{center}
\begin{Large}
{\bf
HBT Effect Based on a Hydrodynamical Model
}
\\  \vspace{1cm}
\end{Large}
\begin{large}
Kenji {\sc Morita}, Shin {\sc Muroya}$^*$ and Hiroki {\sc Nakamura}
\\
\vspace{.5cm}
{\it
Department of Physics, Waseda University, Tokyo, 169 Japan \\
$^*$Tokuyama Women's College, Tokuyama City, Yamaguchi, 745 Japan
}
\end{large}
\end{center}
\vspace{1cm}

\begin{abstract}
We investigate HBT radii based on the numerical solutions
of the hydrodynamical model which are so tuned as to 
reproduce the recent experimental data at the 
CERN SPS. Comparing the sizes of freeze-out hypersurface 
with HBT radii, 
we discuss dynamical effect of the systematic flow on the 
apparent HBT radii. 
Finally  we compare HBT radii of the QGP phase 
transition model with those of the hadron gas model 
without phase transition. 
\end{abstract}

Hanbury-Brown Twiss (HBT) effect\cite{rf:org} is a well known quantum effect 
which enables us 
to estimate the source size through the two-body correlation of the 
emitted particles. 
In the high energy nuclear collision, the correlation experiments of 
pions, kaons, 
etc are promising for obtaining the knowledge of the space-time size of the hot 
and dense region during the reactions.
However, in the nuclear collisions, the reaction takes place 
highly dynamically and the particle source is not static.\cite{rf:hbt} 
The produced hot fire-ball is expected to expand rapidly 
and to cool down in very short time period.
Hence, the meaning of the ``size" of the source (fire-ball) is 
not trivial.

In previous papers, we analyzed numerically (3+1)-dimensional 
hydrodynamical model \cite{rf:hydro} and 
applied it for the recent experimental data of of CERN WA80.
In this paper, based on the numerical solutions we evaluate 
two-particle correlation from which we estimate the HBT radii.
Then we compare these results 
with the size of freeze-out hypersurface.
In particular, we analyze the effect of systematic flow 
of the fluid.  Finally, in order to investigate the signature of
QGP, we compare the HBT radii of the 
phase transition model and those of the hadron gas model without 
phase transition.

An annihilation operator of the particle (e.g. pion) emitted from
a source  $J(x)$ is given as
\begin{equation}
  \aout(\V{k}) = \ain(\V{k}) + i \int\, d^4x
      \frac{1}{\sqrt{(2\pi)^3 \cdot 2\omega_k}} \e^{-i k\cdot x} J(x),
\end{equation}
where $\V{k}$ is momentum of the particle and $\omega_k $ is 
on-shell frequency, $\omega_k = \sqrt{\V{k}^2 + m^2} $. 
The subscripts `out' and `in' correspond to
 out-going field and in-coming field, respectively.
Here we assume $J(x)$ is a c-number source with random phase, i.e.,
\begin{equation}
  J(x) = \overline{J(x)} e^{i \phi(x)},
\end{equation}
where $\phi(x)$ is a random number.   The statistical properties 
of $\phi(x)$ are given by 
\begin{equation}
<e^{i \phi(x)}e^{ -i\phi(y)}>=\delta^4(x-y),
\end{equation}
and Gaussian reduction holds for the higher order correlations.
Then one-particle spectrum is given as,
\begin{eqnarray}
 W(\V{k}) &=& \frac{dN}{d^3\V{k}} = < \left\{ <0_{in}|
      \aout^\dagger(\V{k}) \aout(\V{k}) |0_{in}> \right\}> \nonumber \\
      &=& \int d^4x \frac{1}{(2\pi)^3\cdot 2\omega_k}
	  |\overline{J(x)}|^2, \label{eq:onespe}
\end{eqnarray}
where $<...>$ is the average over the random phase and 
$|0_{in}>$ is the vacuum state of the in-coming field.
The c-number source $\overline{J(x)}$ is defined from consistency in 
the hydrodynamical model, i.e., 
one-particle spectrum should be given by the thermal distribution,
\begin{equation}
  \frac{dN}{d^3\V{k}} = \int_{T=T_f} U^{\mu}d\sigma_{\mu}
       \frac{U^{\mu}k_{\mu}}{(2 \pi)^3\omega_k}f(U^{\mu}k_{\mu},T),
\end{equation}
where $U^\mu$ is the four-velocity of the fluid and 
$f(E,T)$ is Bose-Einstein distribution function.
Comparing the above formula with Eq. \ref{eq:onespe}, we can assign 
c-number source function as,
\begin{equation}
 \overline{J(x)} =\left. \sqrt{2U^{\mu}k_{\mu}f(U^{\mu}k_{\mu},T)}
     \right|_{T=T_f}. \label{eq:sf}
\end{equation}
The two-particle distribution is given by,
\begin{equation}
 W(\V{k}_1,\V{k}_2) =
 <\left\{<0_{in}|\aout^\dagger(\V{k}_1) \aout^\dagger (\V{k}_2)
	    \aout(\V{k}_1)\aout(\V{k}_2)|0_{in}>\right\} >,
\end{equation}
and the correlation function which is comparable to
experimental data is obtained after integrating about
the average momenta, $K = \frac{k_1+k_2}{2}$, as 
\begin{equation}
C(q) = \frac{ \int d^3\V{K} W(\V{k}_1,\V{k}_2)}
   {\int d^3\V{K} W(\V{k}_1)W(\V{k}_2)}.
\end{equation} 

Following the conventional manner, 
we define apparent HBT radii, $R_\Side$, $R_\Out$ and $R_\Long$ 
through the Gaussian fitting\cite{rf:par},
\begin{equation}
C_{fit}(\V{q}) = 1+
 \lambda \exp\left\{-\frac{1}{2}\left(R_{\Side}^2q_{\Side}^2
   +R_{\Out}^2q_{\Out}^2+R_{\Long}^2q_{\Long}^2\right)\right\},
\end{equation}
\begin{figure}[t]
\epsfxsize=8cm
\centerline{\epsfbox{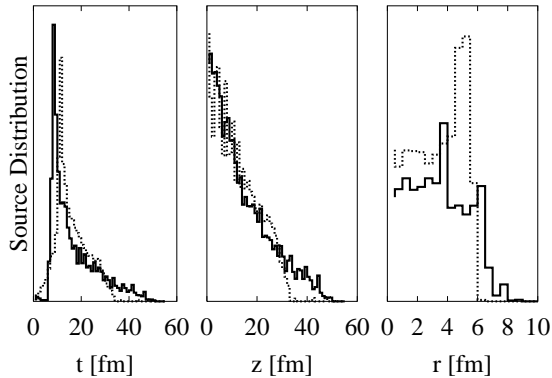}}
\caption{Example of the source distributions as a 
function of $t$, $z$ and $r$, respectively.
The solid line stands for the phase transition model and the dashed line 
stands for the hadron gas model.}
\label{fig:src}
\end{figure}
where $\lambda$ is chaoticity and $q_\Side$, $q_\Out$ and 
$q_\Long$
are the side-ward component, the out-ward component and 
the longitudinal component of $q$, respectively.
Assuming cylindrical symmetry of the Gaussian source function,
and taking the limit of $K \gg q$, we have HBT radii as follows,
\begin{eqnarray}
R_\Side^2 &=& \Delta r^2,\\ 
R_\Out^2  &=& \Delta r^2 + \beta_T^2 \Delta t^2,\\
R_\Long^2 &=& \Delta z^2 + \beta_L^2 \Delta t^2,
\end{eqnarray}
where $\Delta r, \Delta z$ and $\Delta t$ are 
the transverse width, the longitudinal width and 
the temporal width of the source, 
respectively.
$\beta_T$ and $\beta_L$ are the velocities as usual,
$\beta_T =  \frac{\partial E_{\V{K}}}{\partial K_T}$ and 
$\beta_L =  \frac{\partial E_{\V{K}}}{\partial K_L}$.
Though the freeze-out hypersurfaces of our numerical solutions 
are not the Gaussian shape (Fig. \ref{fig:src}),
we evaluate $\Delta r, \Delta z$ and $\Delta t$ as the deviations
of the volume element on the freeze-out hypersurface.

\begin{figure}[t]
\parbox{\halftext}{
\epsfxsize=\halftext
\epsfbox{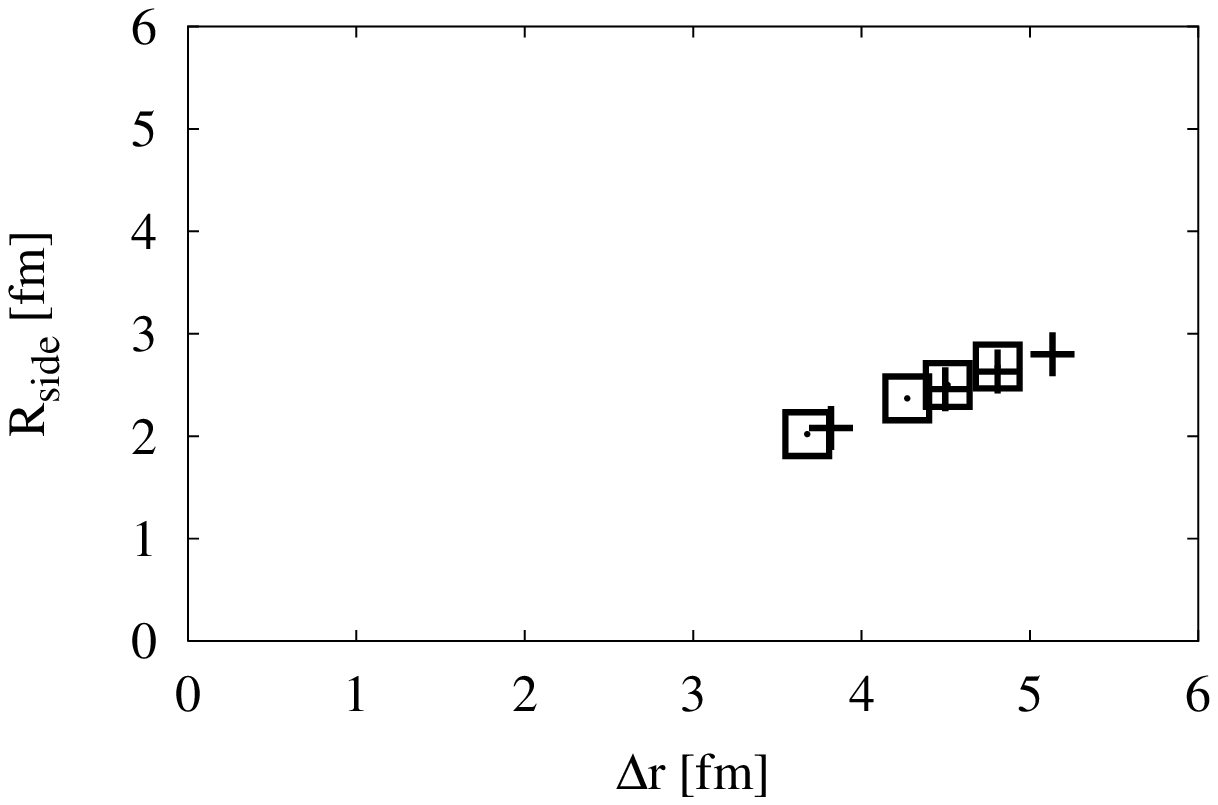}
\caption{$R_{side}$ as a function of $\Delta r$. 
Squares stand for the source without flow and 
crosses stand for the one with flow.}
\label{fig:side}
}
\hspace{8mm}
\parbox{\halftext}{
% \figurebox{6cm}{2cm}
\epsfxsize=\halftext
\epsfbox{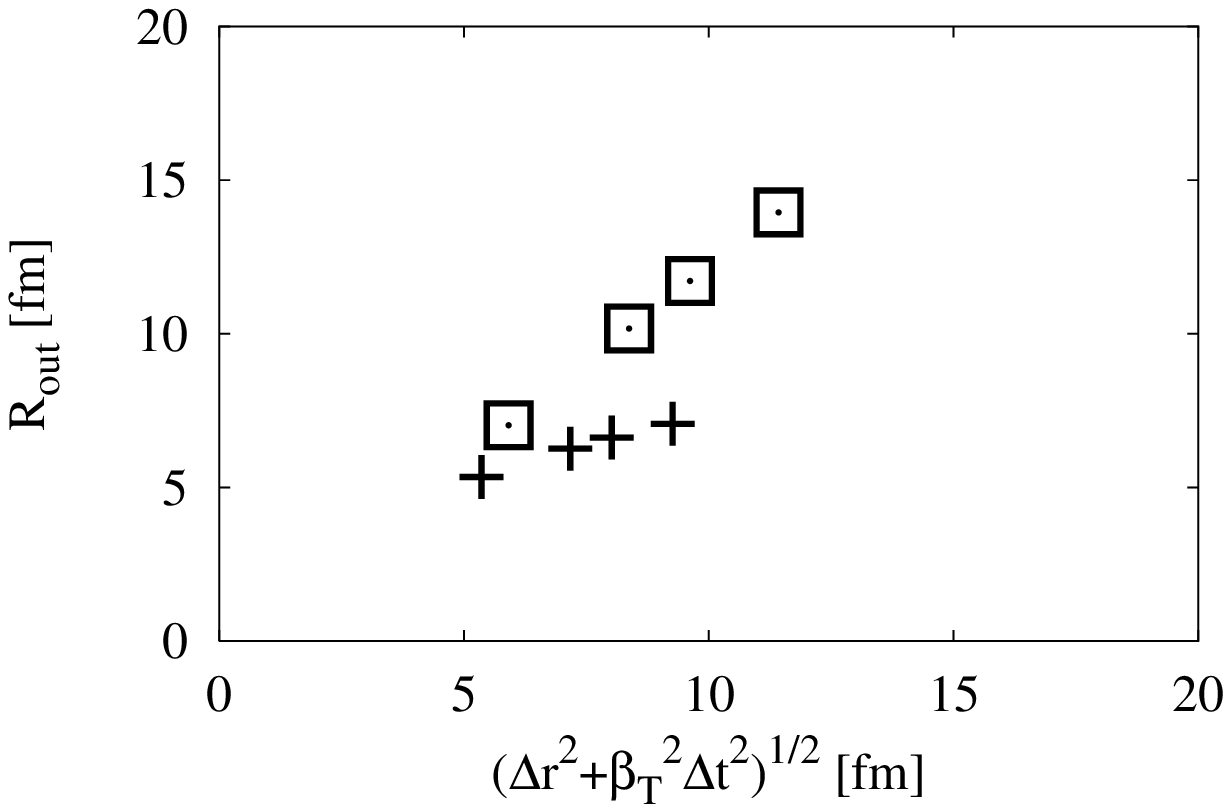}
\caption{$R_{out}$ as a function of $(\Delta r^2+\beta_T^2
\Delta t^2)^\frac{1}{2}$. 
Squares stand for the source without flow and 
crosses stand for the one with flow.}
\label{fig:out}
}
\end{figure}

\begin{figure}[ht]
\parbox{\halftext}{
\epsfxsize=\halftext
\epsfbox{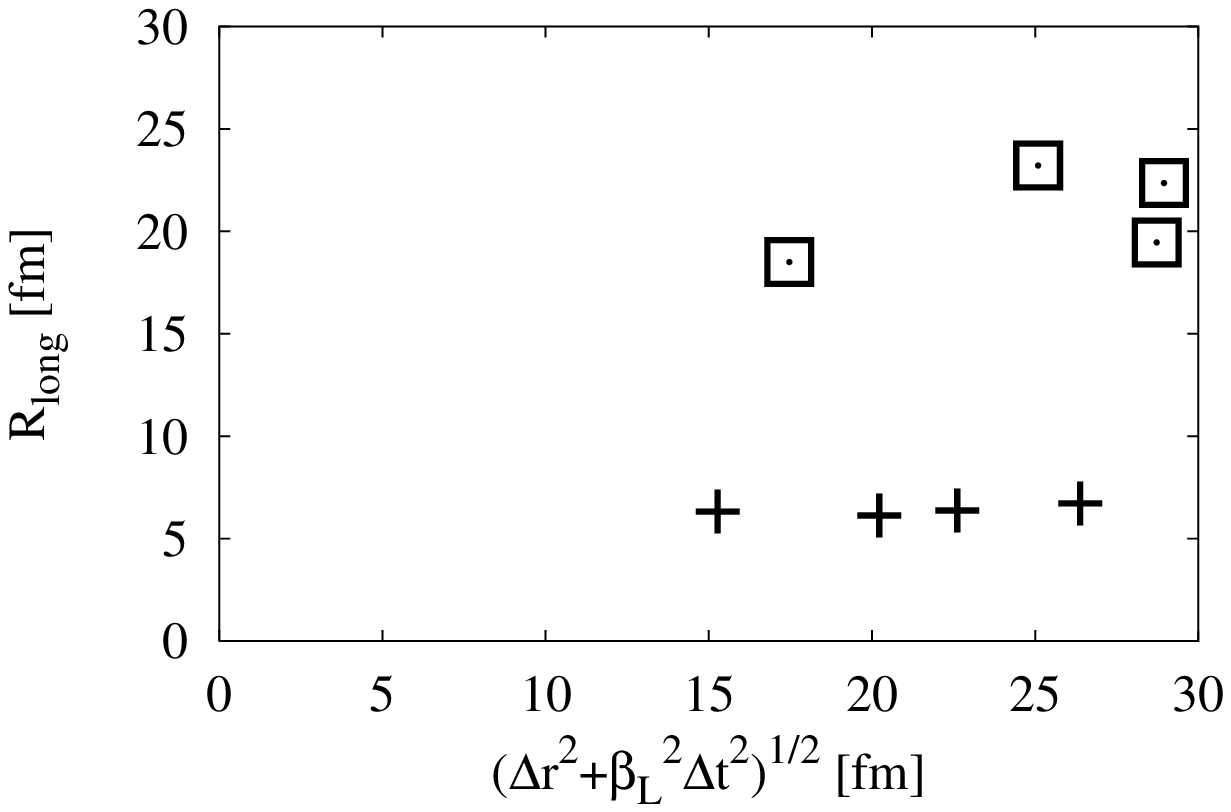}
\caption{$R_{long}$ as a function of $(\Delta z^2+\beta_L^2
\Delta t^2)^\frac{1}{2}$. 
Squares stand for the source without flow and 
crosses stand for the one with flow.}
\label{fig:long}
}
\hspace{8mm}
\parbox{\halftext}{
\epsfxsize=\halftext
\epsfbox{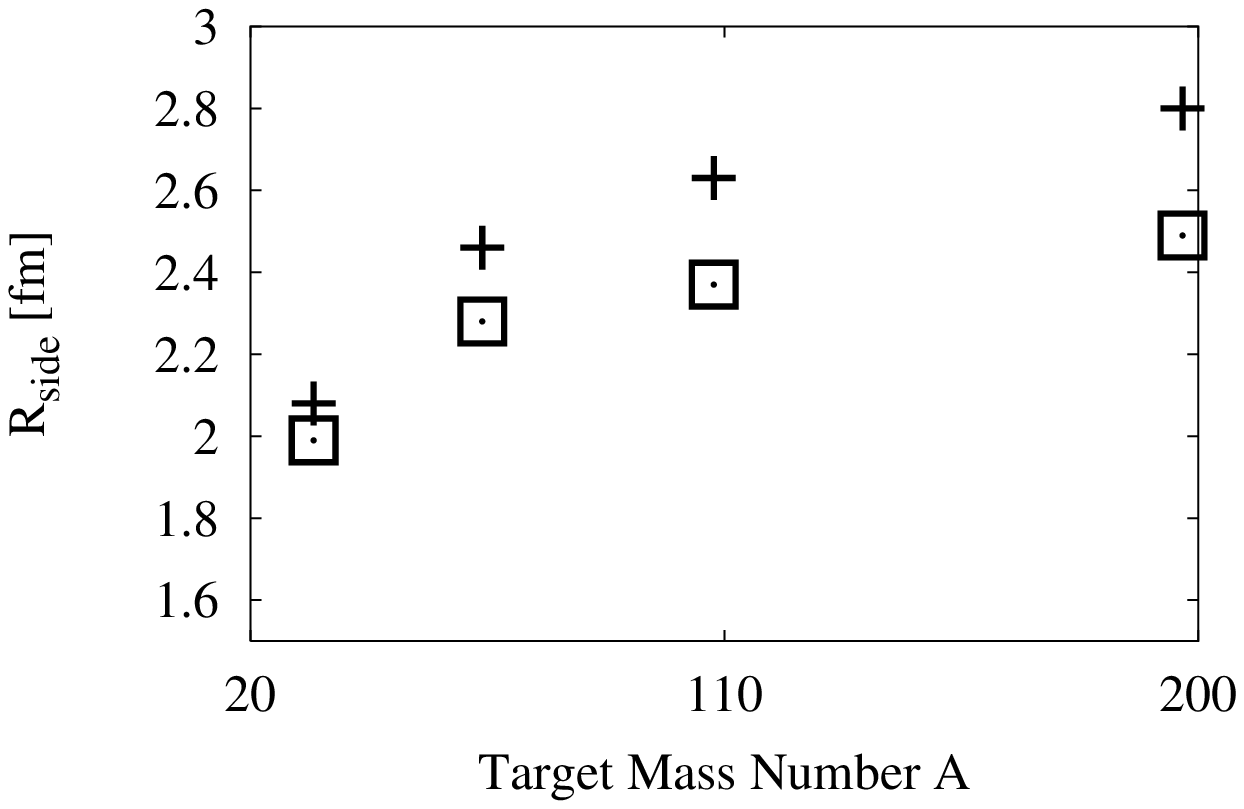}
\caption{$R_{side}$ as a function of target mass number.
Crosses stand for the phase transition model and
squares stand for the hadron gas model.}
\label{fig:ras}
}
\\  
%\end{figure}
%\begin{figure}[ht]
\parbox{\halftext}{
\epsfxsize=\halftext
\epsfbox{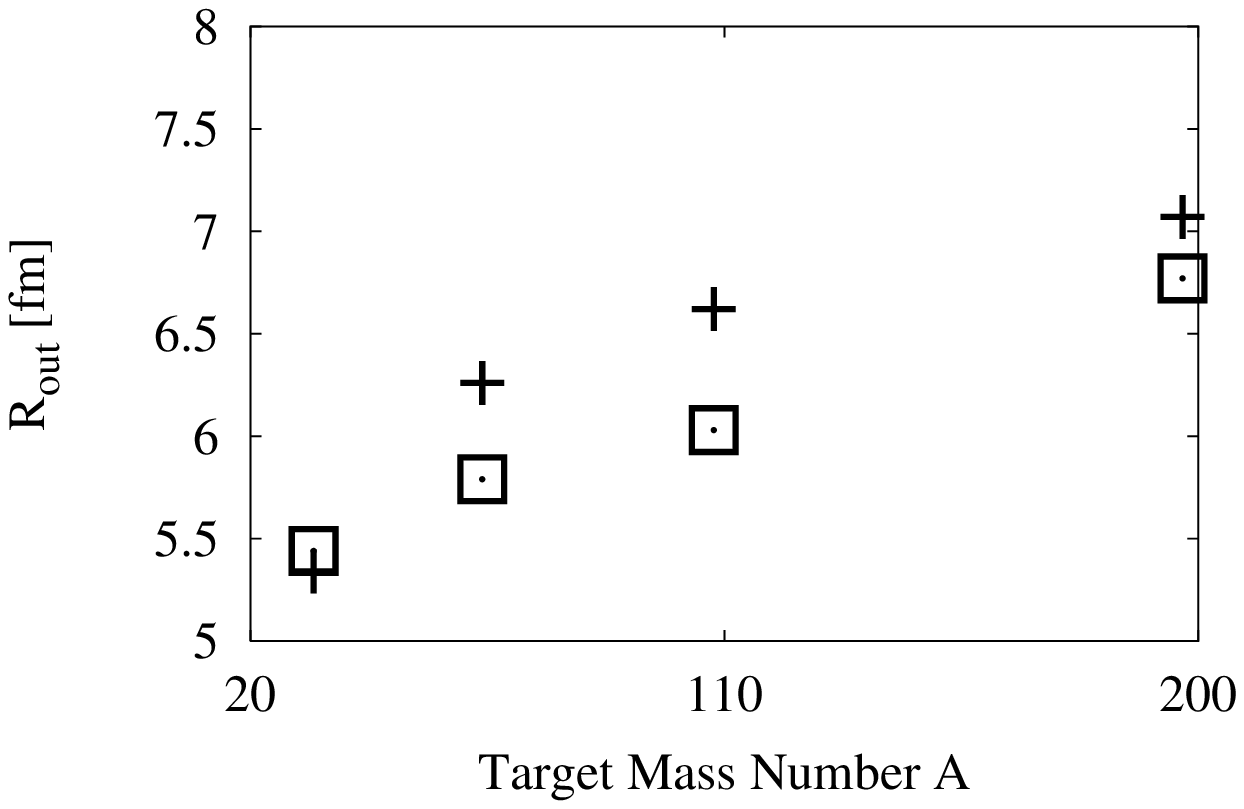}
\caption{$R_{out}$  as a function of target mass number.
Crosses stand for the phase transition model and
squares stand for the hadron gas model.}
\label{fig:rao}
}
\hspace{8mm}
\parbox{\halftext}{
\epsfxsize=\halftext
\epsfbox{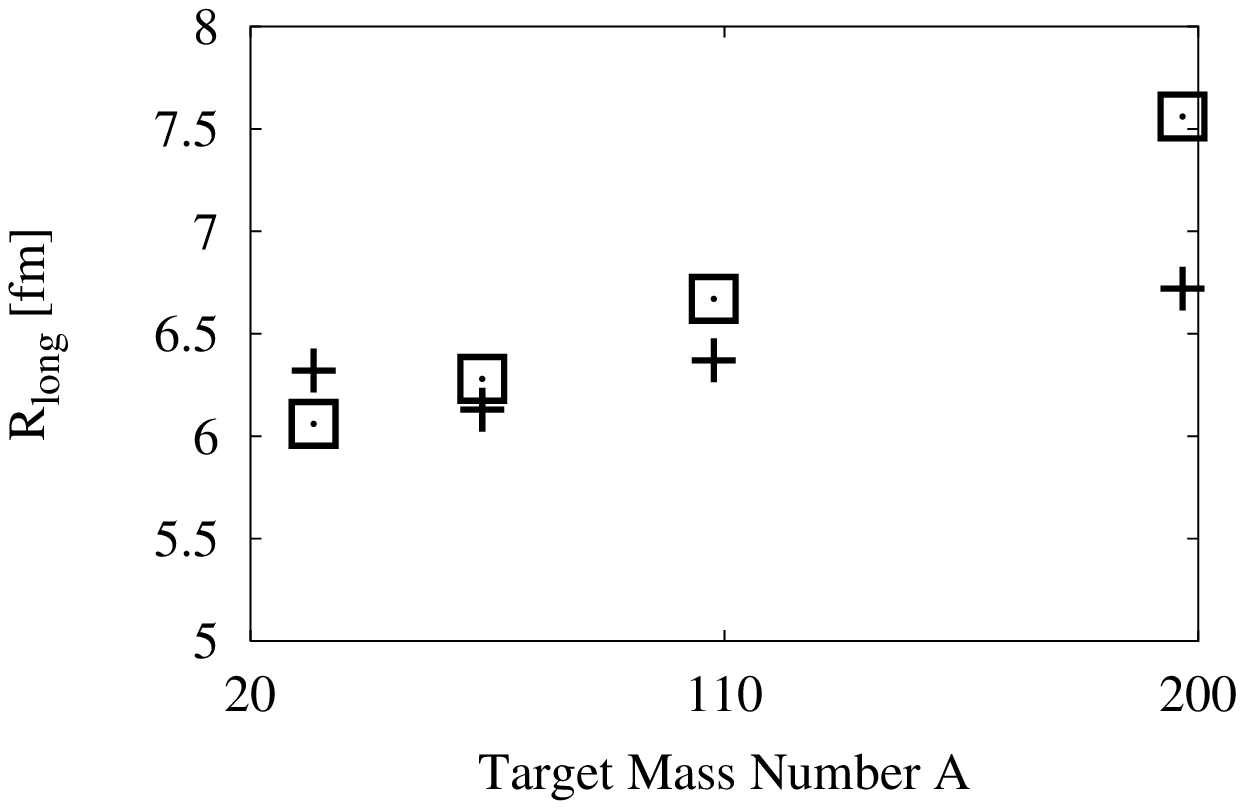}
\caption{$R_{long}$ as a function of target mass number.
Crosses stand for the phase transition model and
squares stand for the hadron gas model.}
\label{fig:ral}
}
\end{figure}

Figures \ref{fig:side}, \ref{fig:out} and \ref{fig:long}
show HBT radii as a function of the hypersurface size.
In order to clarify the effect of the hydrodynamical flow, 
in these figures 
we compared the two sets of data: one is the 
HBT radii as functions of the hypersurface size and the other 
set is the results of static picture.
Data of static picture are evaluated with putting all four 
velocity as $U^\mu = (1,0,0,0)$ by hand.  Four plots in these 
figures correspond to the S+S, S+Cu, S+Ag
and S+Au collisions, from left to right, respectively. 

In figure \ref{fig:side},  $R_\Side$ is proportional to $\Delta r$
 in both data, hence, we may say that $R_\Side$ can 
indicate  $\Delta r$ correctly  
even if the systematic flow exist.  
Because of the non-Gaussian shape of the source, 
the value of $R_\Side$ doesn't equal to the $\Delta r$.
On the other hand, though
$R_\Out$ is proportional to $\sqrt{\Delta r^2 + \beta_T^2 \Delta t^2}$
(Fig. \ref{fig:out}) without systematic flow, 
$R_\Out$, in the case with flow, is not proportional to 
$\sqrt{\Delta r^2 + \beta_T^2 \Delta t^2}$ and becomes smaller than 
it. The systematic flow causes the reduction of 
apparent out-word HBT radius, $R_\Out$. 
When flow exists, $R_\Long$ also become smaller.
However $R_\Long$ is not proportional to the size of source even 
if flow doesn't exist.

We also calculate the two-particle correlation numerically
with use of the hydrodynamical model of the 
hadron gas model without phase transition \cite{rf:hydro}.
Figures \ref{fig:ras}, \ref{fig:rao} and \ref{fig:ral} show
the comparison between the QGP-phase transition model and hadron gas model.
In these figures,  four plots again correspond to 
the S+S, S+Cu, S+Ag
and S+Au collisions, from left to right, respectively. 
The hadron gas model gives larger $R_\Side$ than 
the phase transition model, because $\Delta r$ of the hadron 
gas model is larger than that of the phase transition model 
as shown in Fig. \ref{fig:src}.
In the cases of $R_\Out$ and $R_\Long$, the results of these models 
seem to be hardly distinguishable. 

In this paper,  we compared HBT radii with the size of source calculated 
numerically. $R_\Side$ corresponds to the size of source. 
However, $R_\Out$ and $R_\Long$ become smaller than the size of source
because of the systematic flow. We also compared HBT radii of the 
phase transition model with those of the hadron gas model.
According to our numerical results, it seems very 
difficult to distinguish these models with HBT radii only.

The authors are indebted to Prof.~I.~Ohba and Prof.~H.~Nakazato for their 
fruitful comments.  They also thank to many discussions with 
members of Waseda Univ. High Energy Physics Group.

\end{document}